\documentclass[usenatbib,usegraphicx,letterpaper]{mn2e}

\usepackage{amssymb}
\usepackage{epsfig}
\usepackage{amsmath}
\usepackage{xcolor}
\usepackage[utf8]{inputenc}

\usepackage{hyperref}

\newcommand{\gsam}{{\tt GalSampler}}

\newcommand{\mnras}{{\em MNRAS}}
\newcommand{\apj}{{\em ApJ}}
\newcommand{\apjl}{{\em ApJ Letters}}
\newcommand{\apjs}{{\em ApJ Supplements}}
\newcommand{\na}{{\em New Astronomy}}

\newcommand{\fcam}{\mathcal{F}_{\rm CAM}}
\newcommand{\beq}{\begin{equation}}
\newcommand{\eeq}{\end{equation}}
\newcommand{\bit}{\begin{itemize}}
\newcommand{\eit}{\end{itemize}}
\newcommand{\ben}{\begin{enumerate}}
\newcommand{\een}{\end{enumerate}}
\newcommand{\emdash}{\text{--}}

\newcommand{\gprim}{\vec{g}_{\rm prim}}
\newcommand{\gsec}{\vec{g}_{\rm sec}}
\newcommand{\gvec}{\vec{{\rm g}}}
\newcommand{\dvec}{\vec{\delta}}

\newcommand{\grr}[1]{(g-r)_{\rm #1}}
\newcommand{\gr}{g-r}
\newcommand{\ri}{r-i}

\newcommand{\bt}{\rm B/T}
\newcommand{\mhalo}{M_{\rm halo}}
\newcommand{\mstar}{M_{\star}}

\definecolor{hblue}{HTML}{0921CA}

\begin{document}

\title[Cosmological mocks with GalSampler]
{Generating Synthetic Cosmological Data with GalSampler}

\author[Hearin et al.\ (LSST~DESC)]{Andrew Hearin$^{1}$\thanks{Contact e-mail: \href{mailto:ahearin@anl.gov}{ahearin@anl.gov}}, Danila Korytov$^{1,2}$, Eve Kovacs$^{1}$, Andrew Benson$^{3}$, Han Aung$^{4}$, \and Christopher Bradshaw$^{5}$, Duncan Campbell$^{6}$ \and(The LSST Dark Energy Science Collaboration)
\vspace{0.25cm}
\\
$^{1}$Argonne National Laboratory, Lemont, IL 60439, USA \\
$^{2}$Department of Physics, University of Chicago, Chicago, IL 60637, USA \\
$^{3}$Carnegie Observatories, 813 Santa Barbara Street, Pasadena, CA 91101 \\
$^{4}$Department of Physics, Yale University, P.O. Box 208120, New Haven, CT 06520, USA \\
$^{5}$Department of Astronomy and Astrophysics, University of California, Santa Cruz, 1156 High Street, Santa Cruz, CA 95064 USA \\
$^{6}$McWilliams Center for Cosmology and Department of Physics, Carnegie Mellon University, Pittsburgh, PA 15213, USA
}

\maketitle

\begin{abstract}
As part of the effort to meet the needs of the Large Synoptic Survey Telescope Dark Energy Science Collaboration (LSST DESC) for accurate, realistically complex mock galaxy catalogs, we have developed {\gsam}, an open-source python package that assists in generating large volumes of synthetic cosmological data. The key idea behind {\gsam} is to recast hydrodynamical simulations and semi-analytic models as physically-motivated galaxy libraries. {\gsam} populates a new, larger-volume halo catalog with galaxies drawn from the baseline library; by using weighted sampling guided by empirical modeling techniques, {\gsam} inherits statistical accuracy from the empirical model and physically-motivated complexity from the baseline library. We have recently used {\gsam} to produce the cosmoDC2 extragalactic catalog made for the LSST DESC Data Challenge 2. Using cosmoDC2 as a guiding example, we outline how {\gsam} can continue to support ongoing and near-future galaxy surveys such as the Dark Energy Survey (DES), the Dark Energy Spectroscopic Instrument (DESI), WFIRST, and Euclid.
\end{abstract}


\section{Introduction}
\label{sec:intro}

In order to prepare for the arrival of data from the Large Synoptic Survey Telescope\footnote{\url{http://www.lsst.org}} \citep[LSST,][]{ivezic_etal08,lsst_science_book,lsst_srm}, the Dark Energy Science Collaboration\footnote{\url{http://www.lsst-desc.org}} (DESC) has begun a series of three data challenges of increasing scope and complexity, referred to as DC1, DC2, and DC3. A major component of these challenges is the production of mock catalogs containing galaxies with increasingly complex sets of properties. Analysis working groups within DESC use the DC catalogs to provide a controlled environment for testing systematic effects that will be present in LSST image analysis, and so it is critical that these mocks resemble LSST data as closely as possible.

Due to the diversity of science that the LSST enables, the science driving these data challenges benefits from mocks containing galaxies with a large variety of properties. Example attribute requests from various analysis working groups include broadband flux through filters of several different surveys; spectral energy distributions (SEDs) for photo-z analysis; internal morphological structure such as bulge-disk decomposition (with separate SEDs and fluxes per component), half-light radius, radial profiles and axis ratios; gravitational shear, convergence and magnification; black hole mass and accretion rate. These scientific needs add up: each galaxy in the cosmoDC2 mock submitted to DESC for DC2 analysis contains $\mathcal{O}(500)$ properties.

The cosmological simulations underlying these catalogs are necessarily very large. Currently, hydrodynamical simulations of the necessary volume, resolution, and accuracy are prohibitively expensive, and so gravity-only simulations must be combined with a galaxy model to produce mock catalogs of the required size and characteristics. The DC2 volume requirement also places restrictions on the performance of the method used to populate the simulation with galaxies, which cannot be too expensive as many iterations will be required in order to produce a mock that meets the specified validation criteria.

The evolving nature of a survey's synthetic data requirements plays a critical and largely overlooked role in the challenge of generating mock catalogs for large collaborations. As a survey progresses, changes to the requirements happen for a variety of reasons: additional scientists join the collaboration and bring new expertise that informs the criteria; contemporaneous surveys release new data or measurements; alternative analysis techniques naturally arise and commonly require alternative mocks with features beyond those that were initially planned. The subsequent adjustments to the synthetic catalog may be as simple as modifications to parameter values, or may require introducing entirely new features into the model. This evolving nature is rather fundamental to the operating mode of the collaborations currently conducting large-scale galaxy surveys. We stress that this evolving nature precludes the possibility of a one-time-only ``hero" computation used to calibrate the model. Instead, these practical considerations imply a need for techniques that are quickly adaptable and extensible to new requirements; this efficiency is challenging to achieve due to the simultaneous need for a large variety of galaxy properties with multi-dimensional correlations.

To generate mock catalogs with sufficient complexity, numerous groups use semi-analytic models of galaxy formation (SAMs) grafted into gravity-only $N$-body simulations \citep[e.g.,][]{kauffmann_etal99,kauffmann_haehnelt00,croton_etal06,bower_etal06,delucia_blaizot07,somerville_etal08,guo_etal11,benson_galaxy_2012,overzier_etal13,lacey_etal16,henriques_etal17,lagos_etal18}. In the SAM approach, galaxy properties are determined by solving a set of coupled differential equations modeling a range of (baryonic) physical processes that are not present in the simulation, but that are presumed to influence real galaxies. These systems of differential equations are solved on a halo-by-halo basis, and so in mock catalogs that are based directly on SAMs, each synthetic galaxy has a unique SED, morphology, and historical evolution \citep[see][for a recent review, as well as \S\ref{sec:comparison} for further discussion]{somerville_physical_2015}.

While contemporary SAMs naturally produce synthetic galaxies with a high degree of complexity and a large number of attributes, a drawback of this approach is the computational expense of fitting the model parameters \citep[although see][for significant recent progress in this direction]{van_daalen_etal16,henriques_etal17}. This drawback is exacerbated by the evolving nature of the fitting data discussed above. An alternative approach is to generate mock catalogs using data-driven empirical models such as the Halo Occupation Distribution \citep[HOD,][]{berlind_etal03,zheng_etal05}, the Conditional Luminosity Function \citep[CLF,][]{yang_etal03,vdb_etal13}, and abundance matching \citep{kravtsov_etal04,conroy_etal06,behroozi_etal10,moster_etal10}. Models of synthetic galaxies based on empirical methods are computationally orders of magnitude faster than SAMs, but face a challenge that SAMs do not: conventional data-driven approaches only generate galaxies with a highly restricted set of properties. Thus considerable effort is required to transform mocks based on HOD-type methods into synthetic catalogs of galaxies with the required attributes \citep[for a recent review, see][]{wechsler_tinker18}.

In this work, we present a new method for generating synthetic cosmological data that represents a hybrid of the empirical and semi-analytic approaches to the problem. In our method, we begin with a ``baseline" mock catalog generated by a semi-analytic model or hydrodynamical simulation. We then scale up the baseline mock into a cosmological survey-scale simulation, using weighted Monte Carlo sampling to improve the statistical realism of the resulting catalog. While the technique benefits from high-accuracy baseline catalogs, for many applications the parameters of the underlying model need only be coarsely trained.

We outline the basic assumptions made by the {\gsam} technique in \S\ref{sec:fundamentals}, and in \S\ref{sec:scaleup} we discuss our framework for scaling up high-resolution mocks into large-volume simulations. The core techniques for weighted Monte Carlo sampling are presented in \S\ref{sec:matchup}, and in \S\ref{sec:applications} we describe how we applied {\gsam} to generate the cosmoDC2 mock used in LSST DESC DC2. We discuss ongoing work incorporating systematic effects in \S\ref{sec:systematics}, compare our methodology to alternative approaches in \S\ref{sec:comparison}, and conclude in \S\ref{sec:conclusion} with a discussion of future directions.

\section{Core Assumptions}
\label{sec:fundamentals}

A core assumption made by empirical models of large scale structure is that the galaxy-halo connection can be captured by modeling dependencies between a few key physical quantities. We can express this assumption mathematically as follows. Let $\gvec$ refer to the set of all properties exhibited by galaxies in the model universe; we write this vector space as $\gvec=\gprim\times\gsec,$ where $\gprim$ is a small subset of ``primary" galaxy properties with a tight statistical connection to the cosmic density field, $\dvec,$ and $\gsec$ refers to all remaining ``secondary" galaxy properties. Using $P(\gvec\vert\dvec)$ to denote the complete specification of the statistical connection between galaxies and the cosmic density field, we have
\begin{equation}
P(\gvec\vert\dvec)=P(\gprim; \gsec\vert\dvec) = P(\gprim\vert\vec{\delta}) \times P(\gsec\vert\gprim;\vec{\delta}).\nonumber
\end{equation}
Our core assumption is that there exists a decomposition of galaxy properties $\gvec=\gprim\times\gsec$ such that the following approximation holds:
\begin{equation}\label{eq:fundamental_decomposition}
P(\gsec\vert\gprim;\vec{\delta})\approx P(\gsec\vert\gprim).
\end{equation}
Under this assumption, our basic equation for the galaxy-halo connection becomes:
\begin{equation}\label{eq:funeq}
P(\gvec\vert\dvec) \approx P(\gprim\vert\dvec)\times P(\gsec\vert\gprim).
\end{equation}
Equation \ref{eq:funeq} is a statement about the sparsity of the galaxy-halo connection: once a few core galaxy properties are specified, the remaining variables exhibit negligible residual correlation with the density field.
This decomposition considerably simplifies the task of capturing the full complexity of $P(\gvec\vert\dvec),$ with all its nonlinearities and high-dimensional correlations. In generating the cosmoDC2 extragalactic catalog, we leveraged this simplification by empirically modeling the galaxy-halo connection $P(\gprim\vert\dvec)$ in a relatively low-dimensional space, while modeling the quantity $P(\gsec\vert\gprim)$ by sampling galaxies generated by a fine-grained physical model (see \S\ref{sec:applications} for details).

There is good reason to believe that the galaxy-halo connection in the real Universe is indeed relatively sparse. The relation between stellar mass $\mstar$ and halo mass $\mhalo$ exhibits a well-known monotonic scaling relation with tight scatter across a large range in both mass and cosmic time \citep{kravtsov_etal04,conroy_etal06,behroozi_etal10,moster_etal10}. The star formation rate of active galaxies shows a simple power-law scaling with both $\mstar$ \citep[the ``star-forming sequence", e.g.,][]{brinchmann_etal04,noeske_etal07,daddi_etal07,karim_etal11} and $\mhalo$ \citep{moster_etal17,behroozi_etal18}. Galaxy size has a tight and well-measured scaling with $\mstar$ \citep{shen_etal03,huang_etal13,huertas_company_etal13a,lange_etal15,zhang_yang17,shibuya_etal15,huang_etal17}, and a linear scaling with dark matter halo radius across a large dynamic range \citep{kravtsov13,hearin_etal17}. Despite the complexity of galaxy formation physics, the statistical relationships between a wide variety of galaxy properties admit a strikingly simple description.

Of course, the scaling relations employed by empirical models are merely statistical trends that emerge from highly complex physical processes, and so it is natural to expect that the assumption of a sparse galaxy-halo connection must fail at some level. Thus no matter how faithfully $P(\gprim\vert\dvec)$ is captured, there exists some systematic error, $\delta P(\gsec\vert\gprim;\dvec),$ associated with residual correlations between $\gsec$ and $\dvec$ at fixed $\gprim.$ For purposes of generating synthetic data for large imaging surveys, the required tolerance for error on the galaxy-halo connection depends sensitively upon the science target in question. For example, many analyses of image deblending primarily require a synthetic dataset with a broad diversity of physically plausible fields of view; in such cases, there is no critical need for the underlying model to exhibit accurate correlations between black hole mass and large-scale environment. On the other hand, there are also science applications that are enabled by a mock catalog whose luminosity- and color-dependent two-point correlation functions exhibits $\mathcal{O}(1\emdash10\%)$ agreement with observations; mocks for such applications have a much lower tolerance on $\delta P(\gsec\vert\gprim;\dvec),$ as well as a greater challenge in faithfully representing the primary quantities $P(\gprim\vert\dvec).$ In the present work, we limit our scope to a general treatment of \gsam-based techniques for tuning the galaxy-halo connection; for more extensive discussion of the validation requirements driving the application of our methodology to the generation of cosmoDC2, we refer the reader to \citet{korytov_etal19}.

\begin{figure}
\centering
\includegraphics[width=8cm]{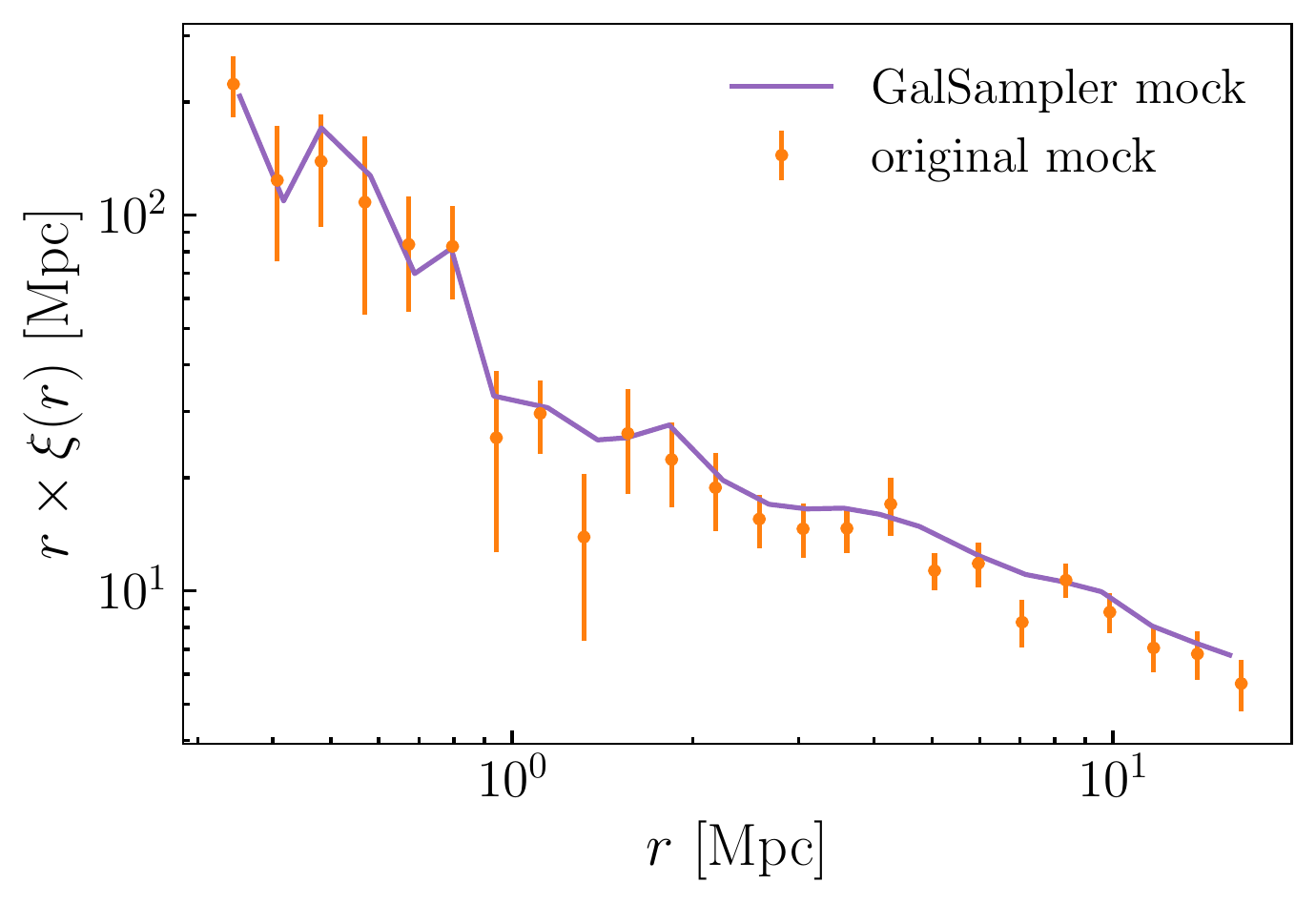}
\caption{The orange points with jackknife error bars show the two-point correlation function of galaxies populating subhalos in the Bolshoi simulation according to best-fit HOD in \citet{leauthaud_etal12}; the purple curve shows the clustering of a \gsam-produced mock galaxy sample populating the larger Multidark simulation. The correlation functions agree at the $\sim20\%$ level on all scales, since {\gsam}  preserves the HOD of the original mock. }\label{fig:elementary_galsampler_recovery}
\end{figure}

\section{Scaling Up High-Resolution Mocks}
\label{sec:scaleup}

In this section we describe techniques for transferring synthetic galaxies populating a modest-volume {\em source halo catalog} into a larger-volume {\em target halo catalog}. The key idea is very simple: for every host halo in the target catalog, identify a counterpart host halo in the source catalog, and map the galaxy content of the source halo into the target halo. A natural way to define the halo--halo correspondence is through mass: for every halo in the target halo catalog, randomly select a halo in the source simulation that has a closely matching mass.\footnote{Note that the halo--halo correspondence is defined in terms of host halos only; subhalos are only relevant to {\gsam} insofar as their positions and properties may be connected to synthetic satellite galaxies. Thus throughout the remainder of this section, the term ``halo" will be understood to mean ``host halo".} Once the halo--halo correspondence has been determined, transferring the galaxy population is straightforward.

Galaxy spatial position and velocity require slightly different treatment, since the Cartesian coordinates of the source and target halo have no connection to one another. If we define the halo-centric distance of each galaxy in the source simulation as $\delta\vec{s}\equiv\vec{s}^{\rm \ 1}_{\rm halo}-\vec{s}^{\rm \ 1}_{\rm gal},$ then we can map the galaxy to the target simulation by assigning it to the position $\vec{s}^{\rm \ 2}_{\rm gal} = \vec{s}^{\rm \ 2}_{\rm halo} + \delta\vec{s}.$ By construction, this will exactly preserve the halo-centric positions of the galaxies.

We demonstrate how to carry out this calculation with a simple, nearest-neighbor-based algorithm in Figure \ref{fig:elementary_galsampler_recovery}, in which we show the two-point clustering of a galaxy sample transferred from one simulation to another. For our source halos, we use a $z=0$ sample of halos identified in the Bolshoi simulation by Rockstar \citep{klypin_etal11,klypin_etal16,behroozi12_rockstar,riebe_etal13,behroozi_etal12b,rodriguez_puebla16_bolplanck}; we populate these halos with source galaxies according to the HOD of $z=0$ galaxies with $M_{\star}>10^{11}M_{\odot},$ using HOD model parameters taken from \citet{leauthaud_etal12}, as implemented in {\tt Halotools} \citep{hearin_etal17}. For target halos, we use a $z=0$ sample of halos identified by Rockstar in the Multidark simulation, using {\gsam} to populate these halos with source galaxies. Bolshoi has a box size of $L_{\rm box}=250\ {\rm Mpc/h}$, while Multidark has a box length of $L_{\rm box}=1\ {\rm Gpc/h},$ and so this application of {\gsam} effectively scales up the simulated volume by a factor of 64. The two curves in Fig.~\ref{fig:elementary_galsampler_recovery} compare the two-point correlation function of the two galaxy samples, which agree at the $10-30\%$ level on all scales. While it has been known for many years that correlation functions are modified by at least this amount when shuffling semi-analytic galaxies amongst halos of a similar mass \citep{croton_etal07}, achieving this level of accuracy was not a validation requirements for cosmoDC2, and so here we restrict discussion to {\gsam} applications based on halo mass alone; further tuning of the two-point functions can be accomplished using the methods discussed in \S\ref{sec:matchup}.

\section{Adjusting the Galaxy-Halo Connection}
\label{sec:matchup}

In \S\ref{sec:scaleup} we described how to transfer a galaxy catalog from one simulation to another while attempting to preserve the complexity of the original catalog. In this section we present two examples demonstrating weighted sampling techniques that can be used to improve the fidelity of an existing mock catalog while retaining as many of its physical features as possible. In the first example discussed in \S\ref{subsec:rescaling}, the modifications to the existing mock are determined exclusively by changes to a single broadband color; the methods shown in \S\ref{subsec:matchup} can be used in more complicated cases where modifications are required of multiple dimensions simultaneously.

\subsection{Weighted sampling with Conditional Abundance Matching}
\label{subsec:rescaling}

Achieving statistically accurate distributions of broadband color across redshift is one of the most challenging aspects of creating synthetic galaxy catalogs. In this section, we demonstrate how to improve the distribution of colors of an existing mock using weighted Monte Carlo sampling.

\begin{figure*}
\centering
\includegraphics[width=12cm]{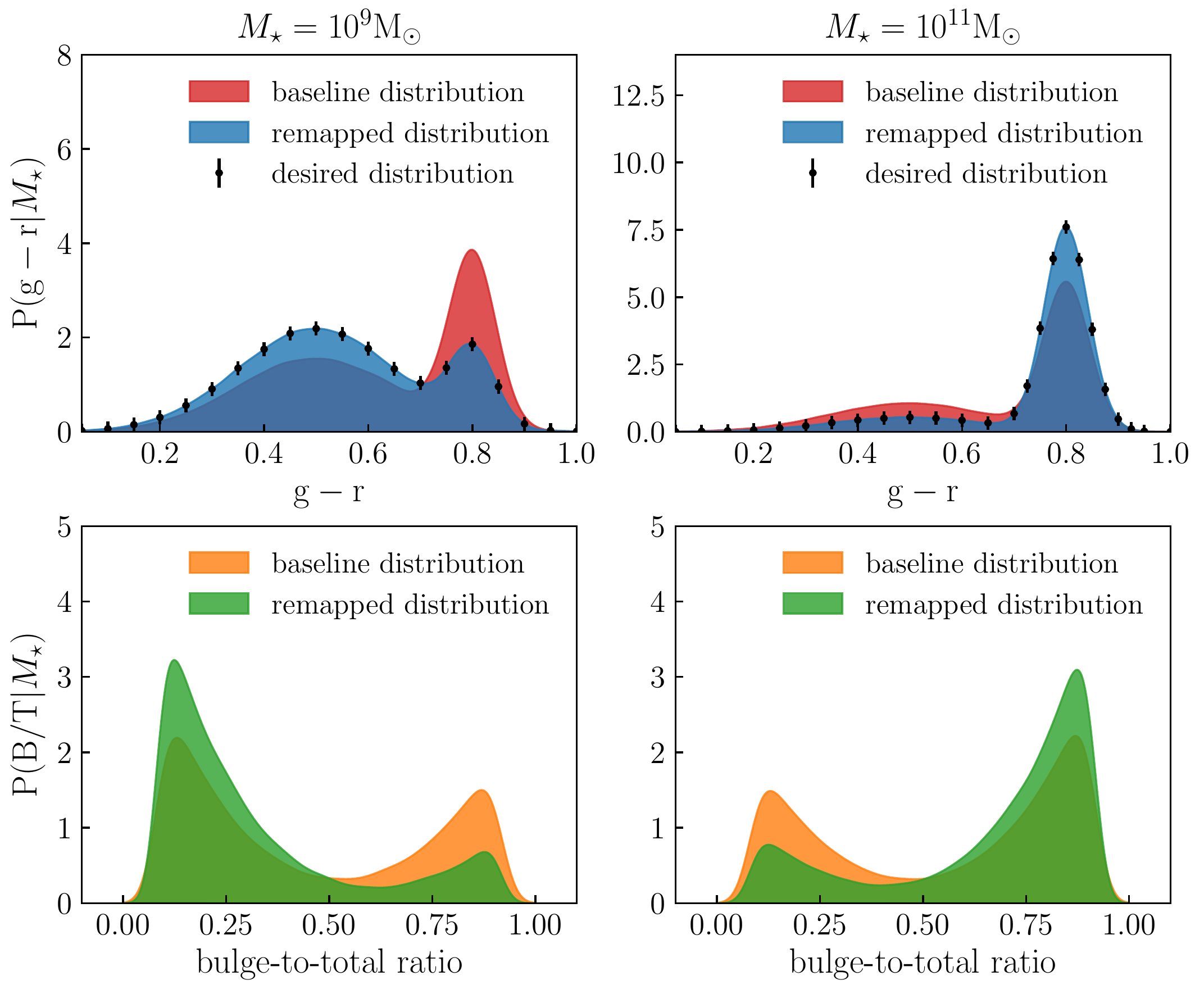}
\caption{
Demonstration of the methods described in \ref{subsec:rescaling}, in which we use conditional abundance matching (CAM) to modify the distribution of $\gr$ color. In the top panels, we compare the $\gr$ distribution of the baseline library (red PDF) to the desired distribution of colors (black points with Poisson error bars). To create the remapped distribution, we resample the galaxies in the baseline mock so that the resulting $\gr$ distribution matches the desired color PDF. Left and right panels display results for different stellar masses. The bottom panels show how the resampling technique induces a modification to the distribution of the bulge-to-total stellar mass ratio, ${\rm B/T}.$ Even though CAM exactly preserves the distribution $P(\bt\vert\mstar;\gr),$ the marginal distribution $P(\bt\vert\mstar)$ is transformed due to the covariance of ${\rm B/T}$ with $\gr$ at fixed $\mstar.$
}
\label{fig:cam}
\end{figure*}

We begin by generating a distribution of $\gr$ colors representing a toy model that we wish to modify; this distribution will typically come from the ``baseline" mock that has been only roughly calibrated (e.g., a SAM or hydro simulation). We additionally specify the $\gr$ distribution that we desire of the final mock catalog; this second distribution represents the validation criteria that the mock should satisfy. For these demonstration purposes, we use a double-Gaussian distribution for both the original and desired PDFs, with parameters that vary with stellar mass; the top panels of Figure \ref{fig:cam} illustrate these two models.

We will remap the $\gr$ colors of the baseline mock so as to preserve the rank order of $\gr$ conditioned upon stellar mass. In this way, galaxies in the original mock that are redder/bluer than average for their mass will remain redder/bluer than average, even though the numerical value of $\gr$ is modified to match $P(\grr{desired}\vert M_{\star}).$ The standard technique for accomplishing this task is Conditional Abundance Matching \citep[CAM,][]{masaki_etal13,hearin_watson13}, which in this example defines a non-parametric mapping from $$\mathcal{F}_{\rm CAM}:\ \{M_{\star}; \grr{baseline}\}\rightarrow\{M_{\star}; \grr{desired}\}.$$ Thus  CAM defines a rank-order-preserving map of $\gr;$ this non-parametric map is determined by equating the conditional cumulative distribution functions (CDFs):
\begin{equation}
\label{eq:cam}
{\rm P(<\grr{baseline}\vert M_{\star}) = P(<\grr{desired}\vert M_{\star}}).
\end{equation}
We refer the reader to Section 4.3 of \citet{hearin_etal14} for associated mathematical derivations.

The \href{https://halotools.readthedocs.io/en/latest/api/halotools.empirical_models.conditional_abunmatch.html}{\tt conditional\_abunmatch} function in {\tt Halotools} provides a bin-free algorithm implementing the mapping defined by Eq.~\ref{eq:cam}. The {\tt conditional\_abunmatch} function estimates $P(<\gr\vert M_{\star})$ for each galaxy by calculating each object's $\gr$-rank-order in a sliding window of stellar mass. The top panels of Figure \ref{fig:cam} illustrate the results of this function applied to our double-Gaussian toy example, with the top-left and top-right panels illustrating results for galaxy samples with different stellar mass. The original distribution is shown with the red shaded PDF; black points with error bars show the desired distribution. The blue shaded PDF shows the distribution that results from application of the {\tt conditional\_abunmatch} function to the red histogram; the final distribution is in close statistical agreement with the desired one, and the {\em relative} $g-r$ color of galaxies at fixed $\mstar$ has been preserved.

\subsubsection{Induced higher-dimensional correlations with CAM}
\label{subsec:rescaling_highdim}

The non-parametric mapping of Conditional Abundance Matching, $\mathcal{F}_{\rm CAM},$ is determined by a pair of conditional CDFs of a single variable. However, Eq.~\ref{eq:cam} in fact implies a mapping between the full multi-dimensional space of the problem. Consider, for example, how modifying $P(\gr\vert M_{\star})$ with $\mathcal{F}_{\rm CAM}$ induces a change in the PDF of $\bt,$ the bulge-to-total mass ratio. It is easy to understand at an intuitive level why $P({\rm B/T}\vert M_{\star})$ will be modified under the $\fcam$ transformation: at fixed $\mstar,$ blue galaxies have a more prominent disk relative to red galaxies; thus if $\fcam$ modifies $P(g-r\vert M_{\star})$ to have a bluer population, then the modified population will have more disk-dominated galaxies simply because blue galaxies have become more abundant.

Mathematically, we can understand this intuitive explanation through the definition of parameter marginalization:
\beq
\label{eq:marginalization}
P({\bt}\vert M_{\star}) = \int P(\bt\vert M_{\star}; \gr)\cdot P(\gr\vert M_{\star}){\rm d}(\gr)
\eeq
The mapping $\fcam$ leaves the first factor in the integrand of Eq.~\ref{eq:marginalization} unchanged, but for any nontrivial application of CAM the second factor is modified, inducing a change in $P({\bt}\vert M_{\star}).$ The same will be true for any variable $x$ for which $P(x\vert M_{\star}; g-r)\neq P(x\vert M_{\star})$ over the range where $g-r$ colors are modified.

To facilitate exploration of these induced correlations, the {\tt conditional\_abunmatch} function in {\tt Halotools} supports a simple variation on the CAM technique by offering the option to return the {\em indices} that resample the desired distribution in a way that respects Eq.~\ref{eq:cam}. Returning the resampling indices of the desired distribution, rather than returning the desired values, facilitates straightforward resampling of the additional variables, and guarantees that $P(\bt\vert M_{\star}; g-r)$ is preserved while $P(\gr\vert M_{\star})$ is modified.

To demonstrate how this resampling technique induces changes in the distributions of other variables, we create an additional toy model for the fraction of stellar mass in each galaxy's bulge component, B/T, modeled as a power law with distinct indices for quenched and star-forming galaxies, such that quenched (star-forming) galaxies are bulge- (disk-) dominated. The original B/T distribution of this toy model is shown with the orange PDF in the bottom panels of Figure \ref{fig:cam}; the green PDF shows the B/T distribution that results from the application of $\fcam$ defined by the top panels.

The transformation from the orange to the green PDFs can be simply understood in terms of Eq.~\ref{eq:marginalization}. In the top-left panel, we see that for low-mass galaxies, the original distribution of $g-r$ color is shifted towards bluer colors in the remapped distribution; this preferentially up-weights the selection of disk-dominated galaxies, and so the green PDF in the lower-left panel displays a larger population of galaxies with smaller values of B/T. The right-hand panels illustrate the complementary case: for massive galaxies, the upper-right panel shows that our toy-model application of $\fcam$ modifies the original distribution to have a redder population, which manifests in the more bulge-dominated population displayed by the green PDF in the lower-right panel.

\subsection{Weighted sampling in higher dimensions}
\label{subsec:matchup}

The Conditional Abundance Matching method described in \S\ref{subsec:rescaling} only offers tunable rescaling of a single ``secondary" variable (such as star-formation rate or broadband color) conditioned upon a single ``primary" variable (such as stellar mass). However, conventional CAM offers no recourse for further improvements if the additional, passively-rescaled variables in the problem do not exhibit good statistical agreement with observations (e.g., if the ${\rm B/T}$ distributions shown in green in the bottom panels of Figure \ref{fig:cam} are in poor agreement with data). In this section we describe an alternative methodology for weighted sampling that addresses this shortcoming. In brief, we will first use empirical techniques to generate a synthetic catalog with a small handful of well-constrained  properties, and then for each empirically modeled galaxy, we find a matching counterpart taken from a library of semi-analytic galaxies.

We begin with any mock catalog based on a traditional semi-analytic model satisfying the following characteristics:
\begin{enumerate}
\item Solutions to the differential equations of the SAM produce model galaxies with the necessary attributes;
\item the parameters of the SAM have been at least coarsely trained so that the mock is in reasonable agreement with the summary statistics of interest;\footnote{Insofar as the selected summary statistics constrain the SAM, this helps ensure realism of $P(\gsec\vert\gprim)$; see discussion in \S\ref{sec:fundamentals}.}
\item the galaxies in the mock {\em densely sample} and {\em span the range} of the space of required galaxy properties.
\end{enumerate}

\begin{figure*}
\centering
\includegraphics[width=6cm]{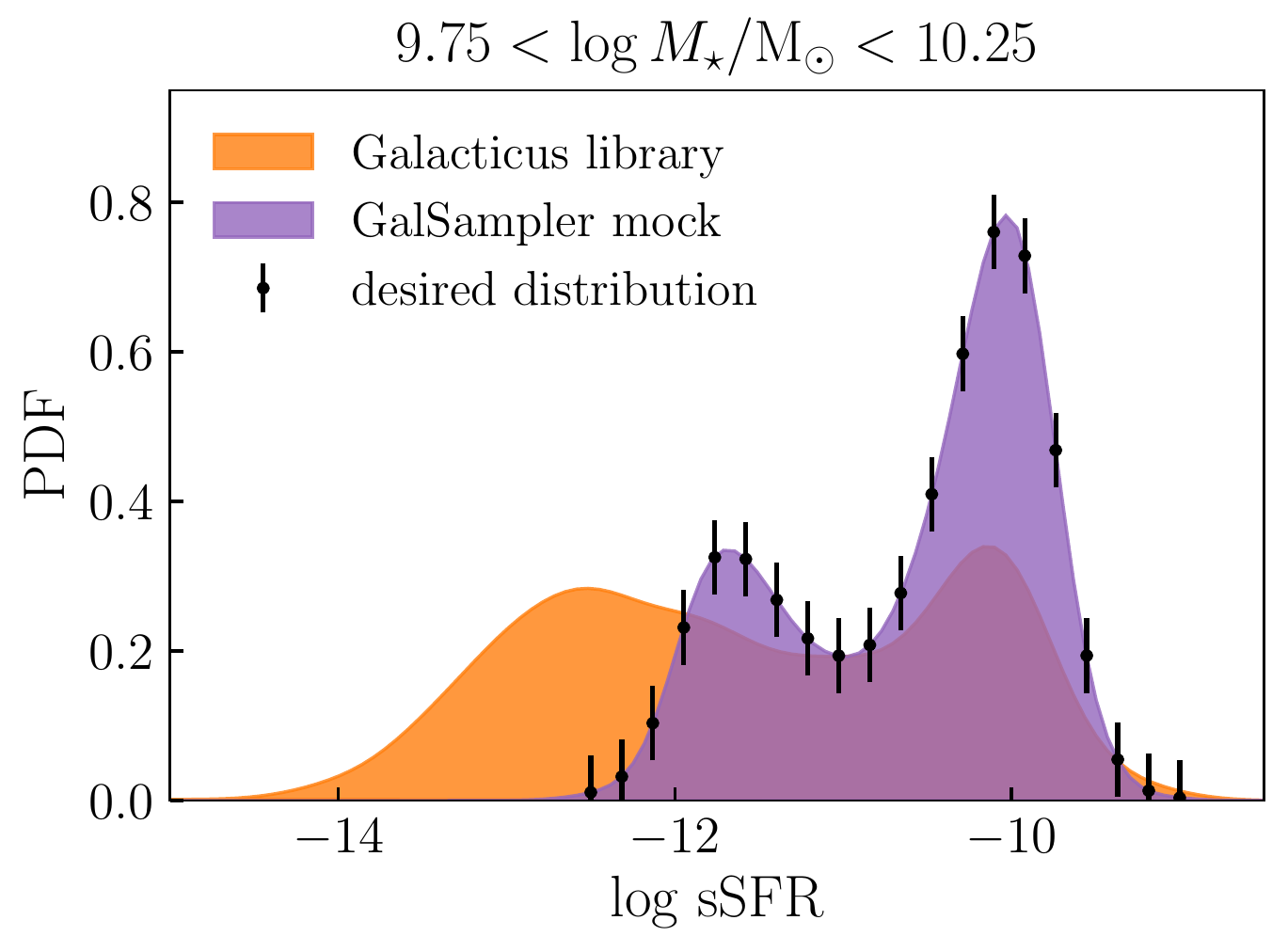}
\includegraphics[width=6cm]{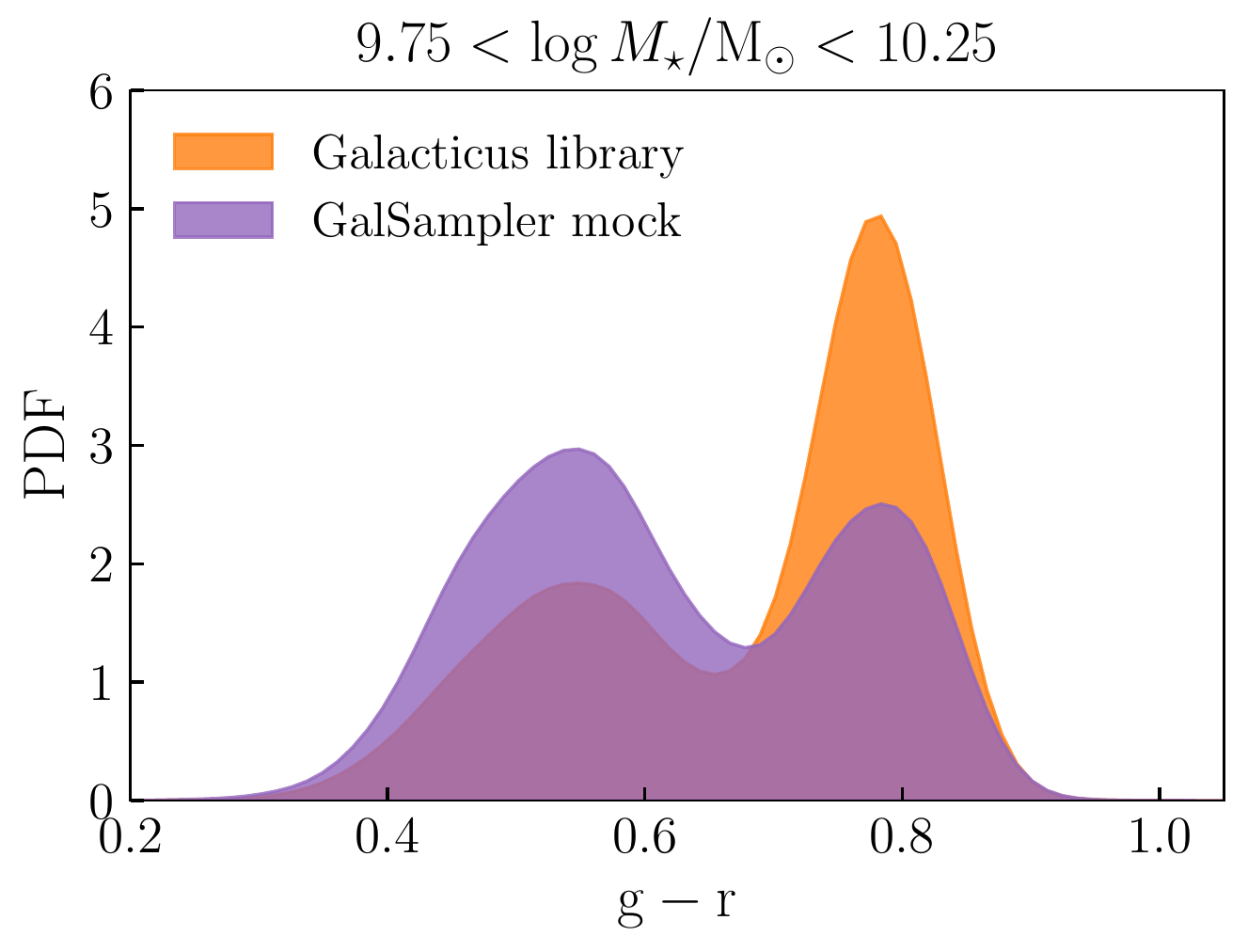}
\includegraphics[width=6cm]{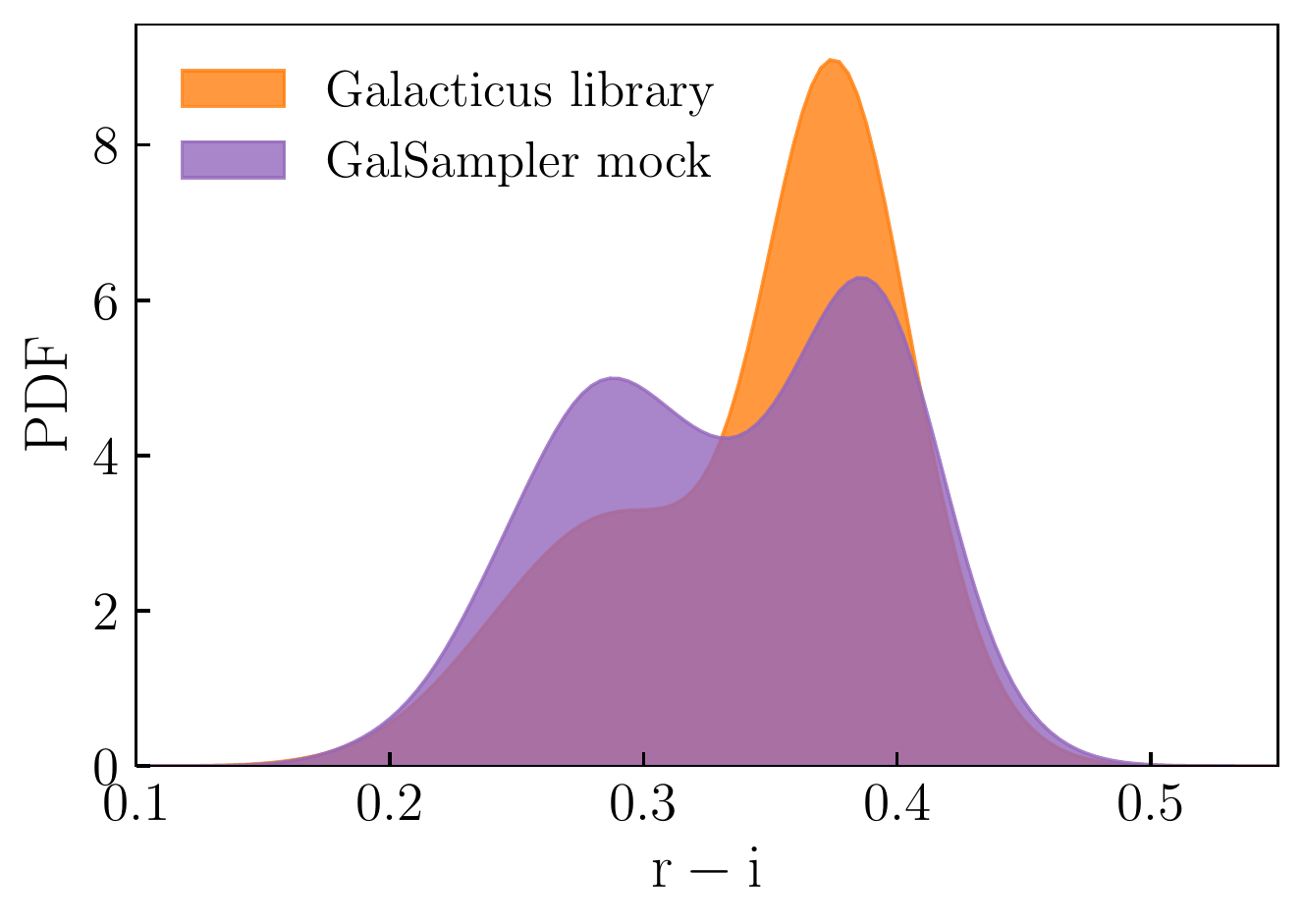}
\includegraphics[width=6cm]{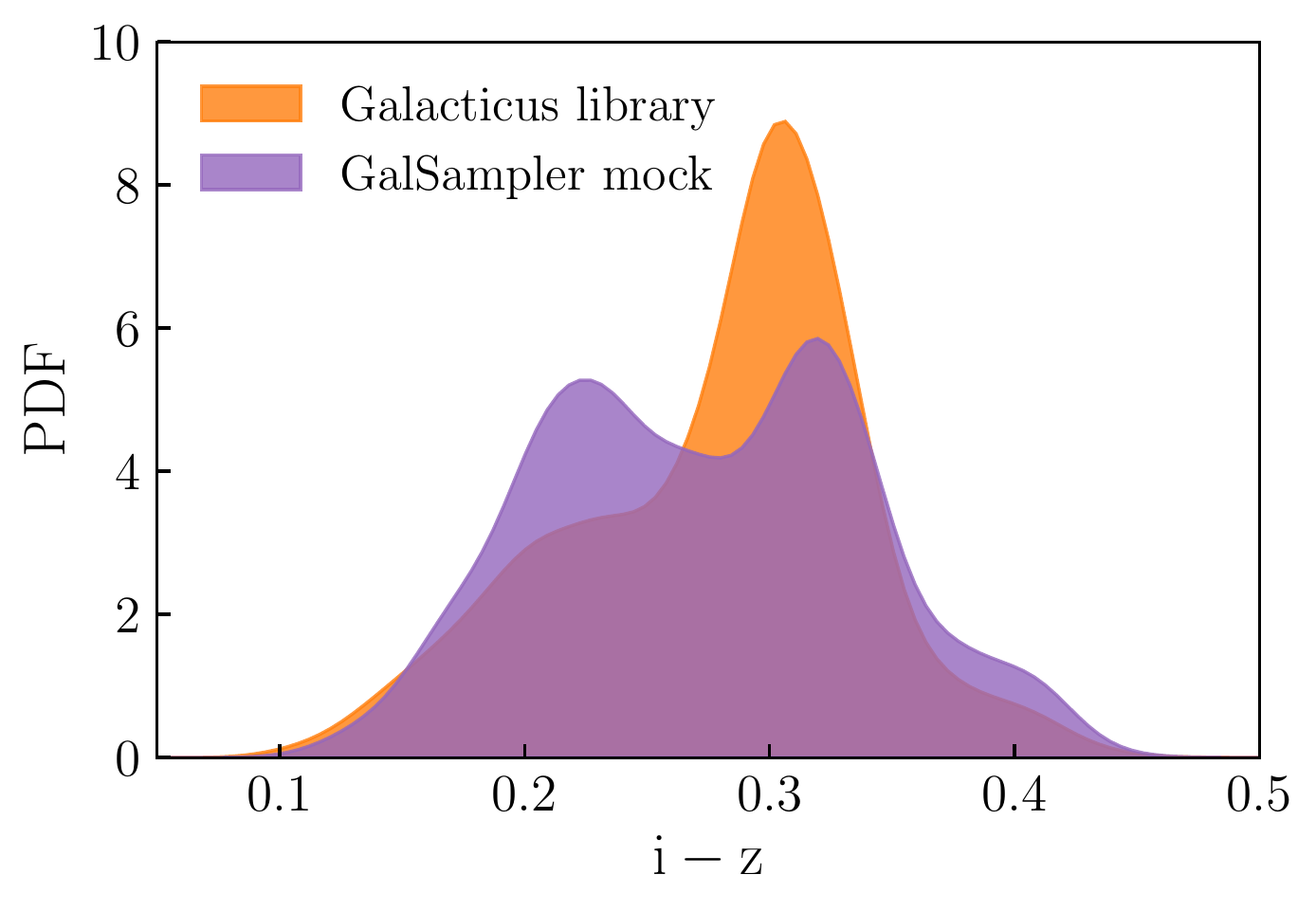}
\caption{Demonstration of the methods described in \S\ref{subsec:matchup}, in which we use multi-dimensional weighted sampling to modify the distribution of star-formation rates. Each panel compares the probability distribution of some property of galaxies in the underlying Galacticus library (orange shaded PDFs) to galaxies in the {\gsam} mock (purple shaded PDFs); for the desired distribution of sSFR, we use the age matching mock (black points with Poisson error bars in the top left panel only); all panels pertain to galaxies in the same stellar mass range, $9.75<\log{\rm M_{\star}/M_{\odot}}<10.25.$ The {top-left panel} compares the distributions of specific star-formation rate, while each of the three remaining panels shows the distribution of broadband color indicated by the horizontal axis label. The desired sSFR distribution in the age matching mock defines the weighted sampling used by {\gsam} to select galaxies from the Galacticus library.}
\label{fig:matchup_colors}
\end{figure*}

\begin{figure*}
\centering
\includegraphics[width=6cm]{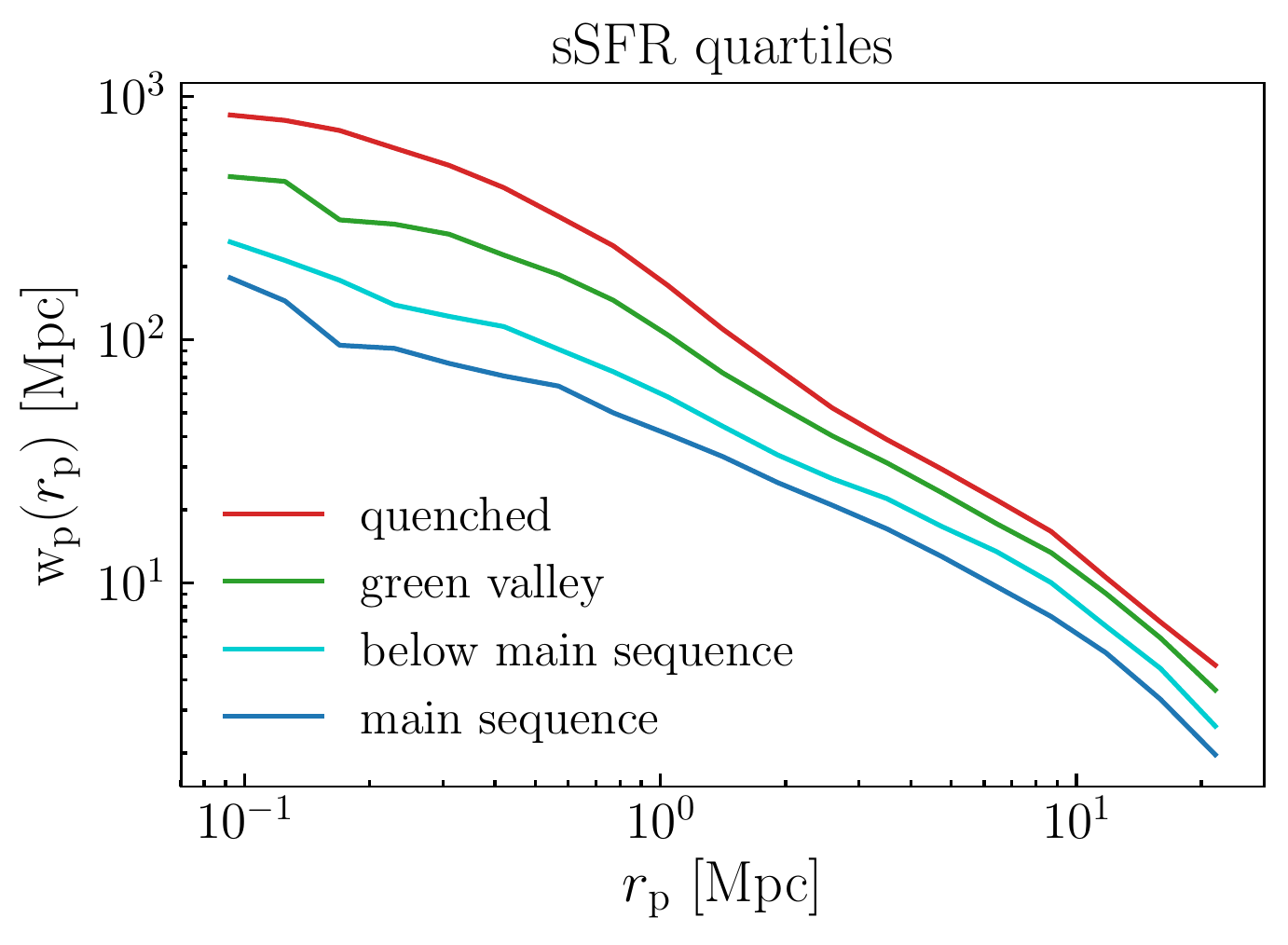}
\includegraphics[width=6cm]{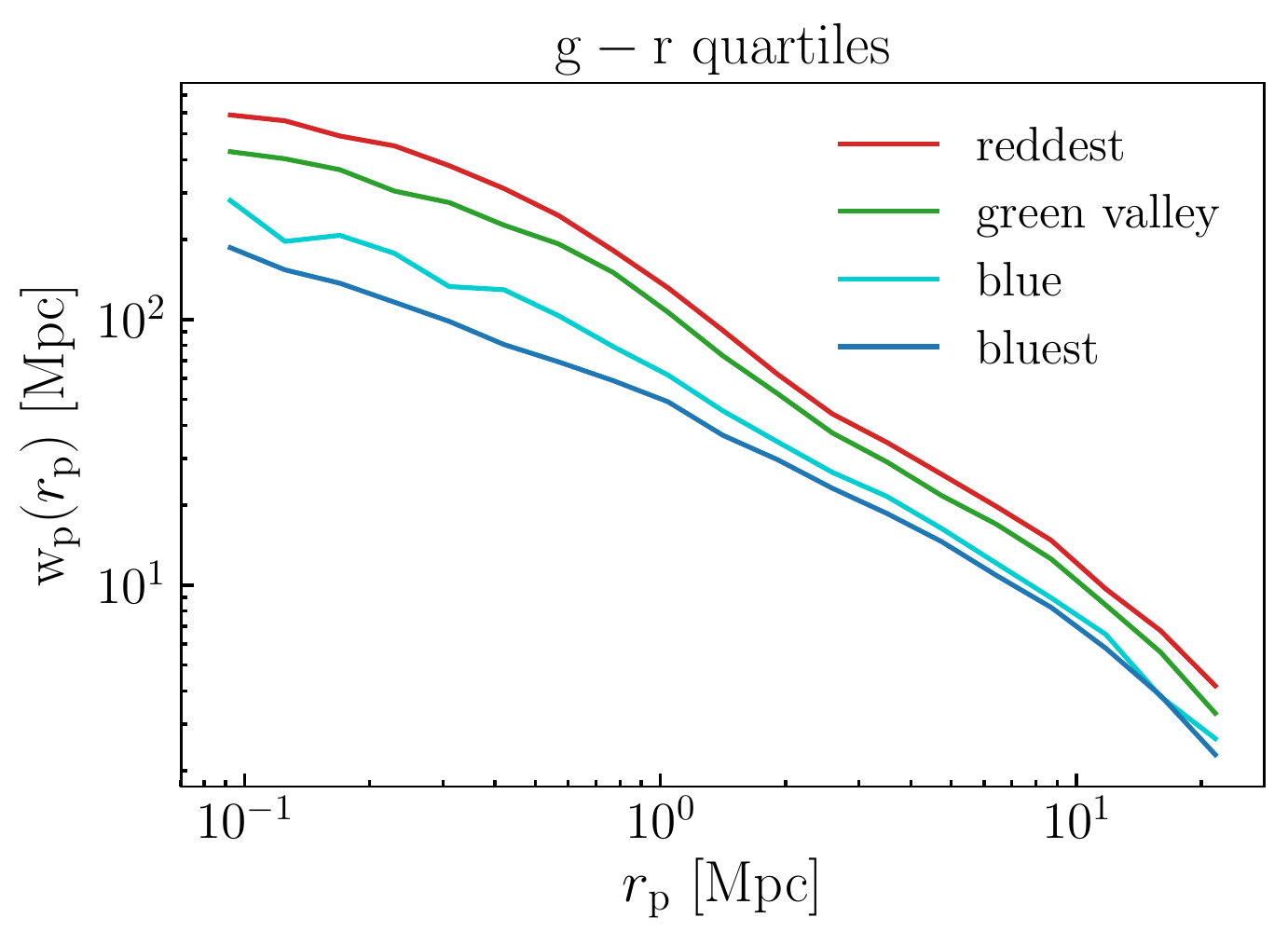}
\includegraphics[width=6cm]{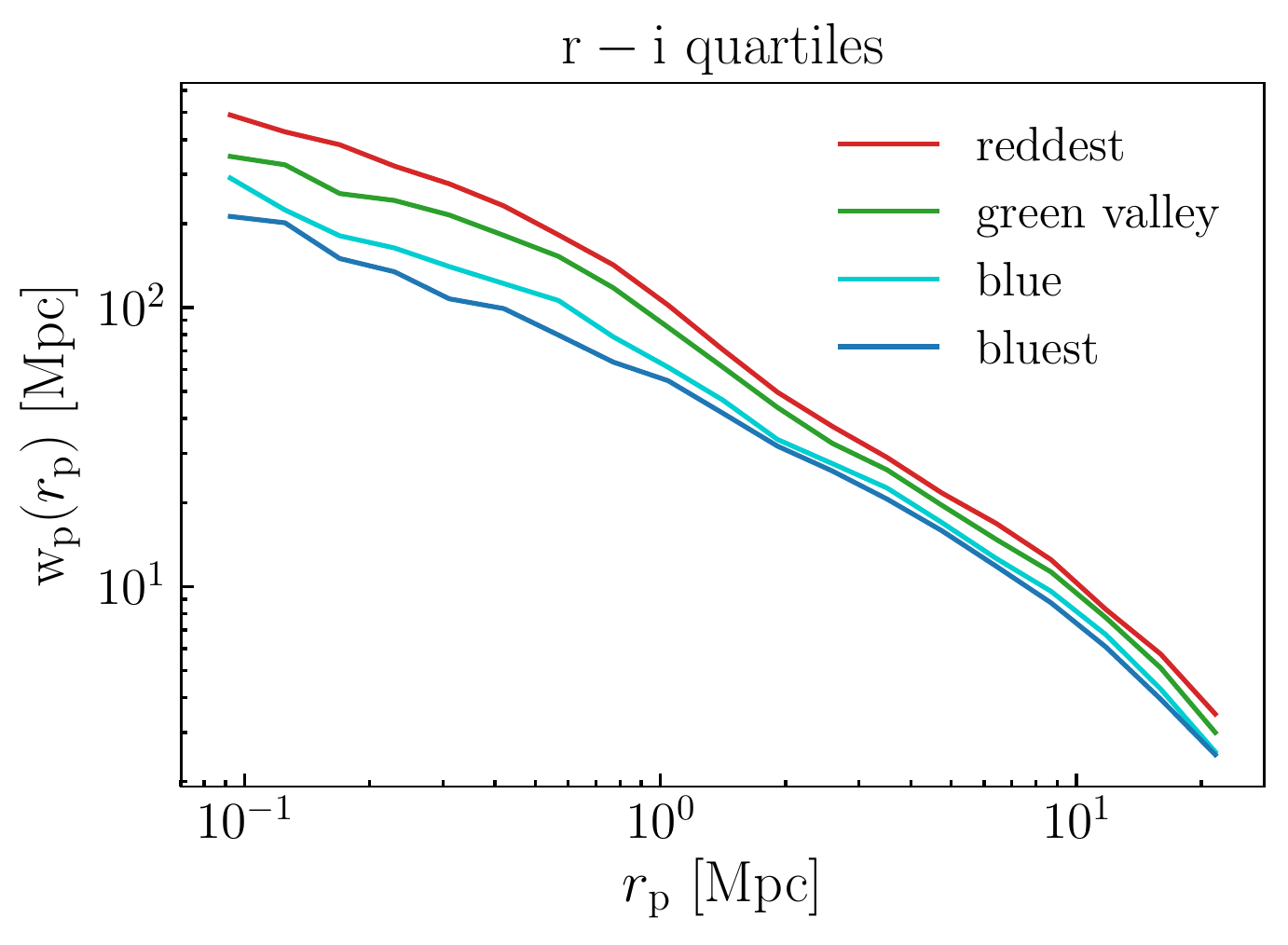}
\includegraphics[width=6cm]{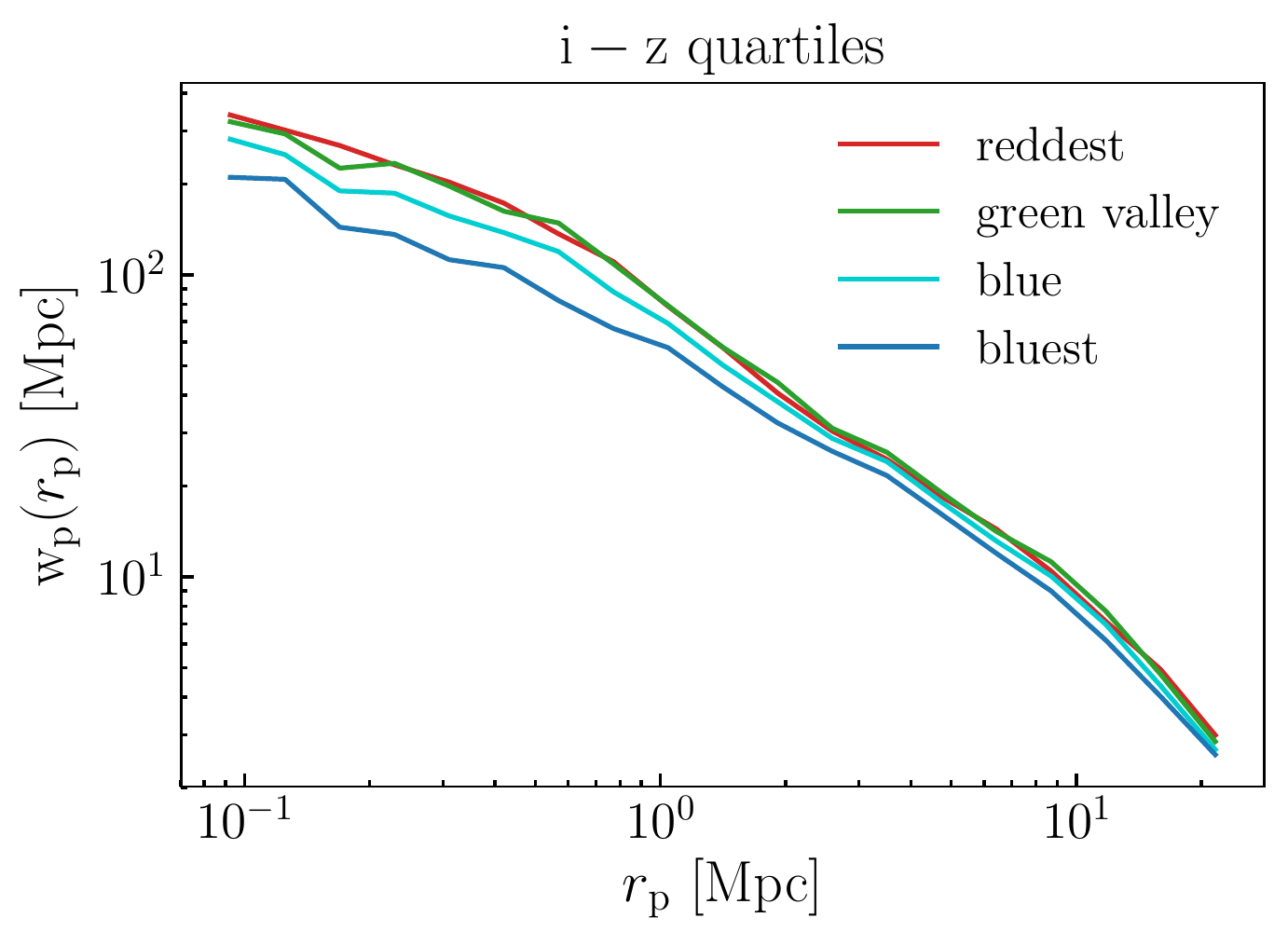}
\caption{Two-point clustering of the {\gsam} mock shown with the orange PDFs in Figure \ref{fig:matchup_colors}. Each panel shows the projected galaxy clustering, $w_{\rm p},$ as a function of projected separation, $r_{\rm p}.$ As in Figure~\ref{fig:matchup_colors}, all panels show results for galaxies in the same stellar mass range, $9.75<\log{\rm M_{\star}/M_{\odot}}<10.25.$ Galaxies are divided into quartiles of the property indicated in the panel's title, and the clustering is shown separately for each quartile. In the top left panel, we show the sSFR-dependent clustering of the {\gsam} mock, which is unchanged from the clustering of the baseline mock based on age matching. The {\gsam} mock inherits broadband color from the Galacticus library, allowing us to calculate color-dependent clustering in the remaining three panels. Because Galacticus predicts that present-day star-formation rate is tightly correlated with each of $\gr,$ $r-i,$ and $i-z,$ then color-dependent clustering in the {\gsam} mock naturally arises via the strong sSFR-dependent clustering in the age matching mock used for tuning.}
\label{fig:matchup_clustering}
\end{figure*}

With such a SAM-based mock catalog in hand, we proceed as follows:
\begin{enumerate}
\item Using empirical modeling techniques, we generate a separate mock catalog with a small handful of core galaxy properties; this mock should faithfully reproduce the highest priority observational measurements used to validate the catalog;
\item for each empirically modeled galaxy, we conduct a nearest-neighbor search in the semi-analytic mock for a galaxy with closely matching core properties;
\item we replace each empirically modeled galaxy with the closely matching SAM galaxy.
\end{enumerate}
The success of this technique relies upon the ability to generate a sufficiently complex and accurate empirical mock, as well as on the ability to find a closely matching SAM galaxy for each empirical one. Under such conditions, the resulting mock will retain all the successes of the empirical model while inheriting the complexity of the SAM. We demonstrate the basic features of this technique with the following toy example.

For our SAM-based mock, we use the publicly available {\tt Galacticus} \citep{benson_galaxy_2012} to create a library of SAM galaxies at $z=0,$ using Alpha-Q as the underlying simulation \citep{korytov_etal19}. Our {\tt Galacticus} mock includes a large variety of model galaxy attributes, such as bulge/disk decomposition with distinct spectral energy distributions in each component. For our empirical model, we use the publicly available mock catalog based on abundance matching and age matching \citep{hearin_watson13}, in which the only modeled galaxy attributes are stellar mass and star-formation rate. For every galaxy in the age matching mock, we use the {\tt cKDTree} implementation in {\tt scipy} to find a {\tt Galacticus} galaxy with a closely matching value of $\{M_{\star},\ {\rm SFR}\}.$ When conducting the search, in practice we use a Euclidean distance metric on variables $x=\log(\mstar)$ and $y=\log({\rm SFR}/\mstar);$ modifying the distance metric allows one to prioritize the fidelity with which different variables will be recovered. Once a nearest neighbor has been found, we then replace each age matching galaxy with its SAM counterpart.

In the top-left panel of Figure~\ref{fig:matchup_colors} we compare the distributions of specific star-formation rates of the empirical and semi-analytic mock; all panels in Fig.~\ref{fig:matchup_colors} display results for $z=0$ galaxies in the same bin of stellar mass, $9.75<\log{\rm M_{\star}}<10.25.$ In this mass range, star-forming galaxies in the {\tt Galacticus} mock are less common relative to age matching, and the {\tt Galacticus} distribution  of specific star formation rate (sSFR) has a much longer tail at the low-end of the sSFR distribution, though the peak of the main sequence occurs at a similar location and level of scatter. More pertinently, the {\tt Galacticus} mock densely spans the range of the sSFR distribution in the age matching mock, and so the sSFR distributions in the {\gsam} and age matching mocks are in close agreement.

The remaining panels of Figure~\ref{fig:matchup_colors} display the distribution of restframe color through SDSS filters: $\gr$ in the top-right panel, $\ri$ in the bottom-left, and $i-z$ in the bottom-right. These broadband colors are not present in the age matching mock, but are inherited from {\tt Galacticus} via the {\gsam} technique. The sSFR distribution of {\tt Galacticus} galaxies is too quiescent relative to the desired distribution in the age matching mock; accordingly, quenched galaxies are preferentially downweighted by {\gsam}, thereby producing distributions of broadband colors that are bluer relative to {\tt Galacticus}.

Figure~\ref{fig:matchup_clustering} shows how the two-point clustering of {\gsam} galaxies depends on star-formation rate and color. Just as in Figure~\ref{fig:matchup_colors}, all panels of Figure~\ref{fig:matchup_clustering} pertain to galaxies in the same bin of stellar mass, $9.75<\log{\rm M_{\star}}<10.25.$ Galaxies in the mass bin are divided into quartiles of sSFR in the top-left panel, and quartiles of each of the three broadband colors in the remaining panels. The sSFR-dependent clustering of the {\gsam} mock is essentially identical to age matching, because every age matching galaxy has a closely matching counterpart in {\tt Galacticus}, and so the sSFR values are essentially unchanged by the {\gsam} technique. Each of the remaining panels shows smooth, monotonic dependence of two-point clustering upon each of the three broadband colors. This feature would be highly impractical to achieve via  standard empirical modeling techniques such as the HOD without a runaway proliferation of parameters. Of course, there is no guarantee that, for example, $(i-z)$-dependent clustering will be in close quantitative agreement with observations; for example, if $i-z$ color in the real Universe is strongly correlated with the cosmic density field at fixed $\{M_{\star},\ {\rm sSFR}\},$ then the resulting {\gsam} mock will not capture such a correlation; we discuss this point further in \S\ref{sec:comparison}. However, a vast array of survey science needs are met by a synthetic catalog with one- and two-point functions that are simply in reasonable absolute agreement with observations, and exhibit correct scaling with all variables in the problem. In \S\ref{sec:applications}, we use the LSST DESC Data Challenge 2 as another example of how to apply the {\gsam} technique to meet the needs of modern galaxy surveys for accurate, complex synthetic catalogs.

\section{Application to LSST DESC Data Challenge 2}
\label{sec:applications}

The previous sections outlined various techniques that can be useful when generating synthetic cosmological data. Each of these techniques played a role in the production of the cosmoDC2 mock catalog for the LSST DESC Data Challenge 2; we give a detailed account of cosmoDC2 in a companion paper to the present work \citep{korytov_etal19}. In this section we will use cosmoDC2 as a worked example to show how the modeling elements discussed above can be tied together in a unified pipeline.

The production pipeline of cosmoDC2 begins with the publicly available mock catalog based on the UniverseMachine empirical model for the galaxy--halo connection \citep{behroozi_etal18}. Each synthetic galaxy in the UniverseMachine mock has an assembly history across cosmic time, including in-situ star formation and mass acquired from mergers, tabulated at every output snapshot of the simulation. The UniverseMachine mock used in cosmoDC2 is based on Rockstar subhalos \citep{behroozi12_rockstar,behroozi_etal12b,rodriguez_puebla16} identified in the MDPl2 simulation \citep{prada_etal12,klypin_etal16}, which was run with the Adaptive-Refinement-Tree (ART) code \citep{kravtsov_etal97}, and has a comoving box size of $1\ {\rm Gpc/h}$ and mass resolution $m_{\rm p}=1.51\times10^{9}M_{\odot}/{\rm h}.$

Using the technique described in \S\ref{sec:scaleup}, we populate host halos in the Outer Rim simulation with UniverseMachine galaxies. Outer Rim was carried out with the Hardware/Hybrid Accelerated Cosmology Code (HACC)~\citep{hacc}, and covers a volume of $3\ {\rm Gpc/h}$ sampled with $10240^3$ tracer particles, leading to a mass resolution of $m_{\rm p}=1.85\times 10^9M_{\odot}/{\rm h}.$ In practice, we populate a lightcone of Outer Rim halos to map our galaxies into the space of $\{{\rm ra, dec, z}\},$ always matching an Outer Rim halo with an MDPl2 halo from the closest-matching snapshot. Upon completing this stage of the pipeline, Outer Rim contains galaxies with statistical properties that approximate those in the original UniverseMachine mock with reasonable fidelity.

Before applying the techniques described in \S\ref{subsec:matchup}, we augment the $M_{\star}$-based galaxy properties with additional modeling of restframe flux through LSST filters $g, r,$ and $i.$ First, we parameterize and calibrate a model for $\langle M_{\rm r}\vert M_{\star}, z\rangle.$ Next, we model restframe colors $\gr$ and $\ri,$ using an independent double-Gaussian for the PDF of each color; for both PDFs, the relative peak heights, peak locations, and scatter vary with $M_{\rm r},\ M_{\rm halo},$ and redshift. We used CAM to introduce correlations between galaxy properties, so that bluer galaxies have larger star-formation rates than redder galaxies for both $\gr$ and $\ri$ colors. For each of these ingredients, we tuned model parameters to meet validation criteria provided by DESC working groups \citep[for details, see][]{korytov_etal19}. At this stage, synthetic galaxies in cosmoDC2 possess the following attributes: $M_{\star},\ M_{\rm g},\ M_{\rm r},\ M_{\rm i},$ stellar assembly history, and redshift.

Finally, as described in \S\ref{subsec:matchup}, for each cosmoDC2 galaxy we conduct a nearest-neighbor search of a library of semi-analytic galaxies, searching the closest-matching snapshot for the SAM galaxy with the smallest Euclidean distance in the space of $\{M_{\rm r},\ \gr,\ \ri\}.$ The SAM library used in cosmoDC2 comes from the {\tt Galacticus} semi-analytic model \citep{benson_galaxy_2012}; the publicly available code was run on a smaller simulation with the same cosmological parameters as those of Outer Rim to produce the output properties requested by DESC working groups. By replacing each empirically modeled galaxy in cosmoDC2 with its nearest-matching {\tt Galacticus} counterpart, we create a synthetic catalog that possesses the large number of attributes required of the mock, while preserving the statistical distributions of the empirical model.

We conclude this section by summarizing the three stages of this modeling pipeline:
\begin{enumerate}
\item Scale up the UniverseMachine mock into the Outer Rim halo lightcone using {\gsam}.
\item Empirically model $\{M_{\rm g}, M_{\rm r}, M_{\rm i}\}$ for every galaxy.
\item Replace each galaxy with a closely matching counterpart taken from the {\tt Galacticus} library.
\end{enumerate}

\section{Incorporating systematic effects}
\label{sec:systematics}

In \S\ref{sec:matchup} we described the core techniques used to produce the cosmoDC2 mock catalog outlined in \S\ref{sec:applications}. However, there are many variations on these methods that can be useful when producing synthetic catalogs, particularly when crafting mocks to study the impact of various systematic errors and modeling assumptions.
Each subsection below outlines a particular way in which we have used weighted Monte Carlo sampling to create a synthetic catalog designed for a specific inquiry into a particular source of systematic error.  Our goal here is merely to illustrate the extensibility of the {\gsam} methodology to a wide range of problems, and so we will relegate a more extensive study of each phenomenon to companion DESC papers using mocks made with these techniques.

\subsection{Assembly bias}
\label{subsubsec:assembias}

One of the key elements of the {\gsam} algorithm is the correspondence between source and target halos. In the application outlined in \S\ref{sec:scaleup}, the halo--halo correspondence was determined exclusively by mass. By construction, this erases any correlation between the galaxy content of a halo and all properties besides the halo's total mass. However, if the galaxy properties in the baseline model exhibit residual correlations with halo properties besides mass, then these additional correlations can be incorporated by including the additional halo properties in the halo-to-halo correspondence. As described above, this can be implemented using the {\tt source\_halo\_index\_selection} function, which supports multi-dimensional halo-to-halo correspondences based on the same nearest-neighbor search algorithm used in the one-dimensional case.

When the source and target halos are defined according to different algorithms, some care must be taken to avoid systematic mismatches that result from inconsistent mass definitions. This concern applies in applications of one-dimensional matching, and can be especially problematic in the multi-dimensional case. To regularize the correspondence in mass-like variables, instead of using mass it is straightforward to match on cumulative number density, ${\rm d}n(<M_{\rm halo})/{\rm d}V;$ the correspondence in a second variable such as halo concentration can be regularized by instead using the conditional cumulative distribution, e.g., $P(<c\vert M_{\rm halo}).$ In this way, we essentially use CAM to set up the halo-to-halo correspondence between the two simulations. Calculating ${\rm d}n(<M_{\rm halo})/{\rm d}V$ can be accomplished by dividing the rank-order of each $x$ and dividing by the simulation volume; the \href{https://halotools.readthedocs.io/en/latest/api/halotools.utils.sliding_conditional_percentile.html}{\tt sliding\_conditional\_percentile} function in {\tt Halotools} can be used to calculate a generic conditional cumulative distribution, $P(<y\vert x);$ while we generally find better recovery of the one- and two-point functions when matching on regularized variables, the need for this numerical technique depends on the application.

\subsection{Satellite anisotropy}
\label{subsubsec:anisotropy}

Since the accretion of satellites onto host halos occurs preferentially along larger-scale filaments, then the properties and/or intra-halo position of satellites in a galaxy catalog may correlate with the larger-scale environment \citep[e.g.,][]{zentner_etal05}. In the applications of {\gsam} outlined in \S\ref{sec:scaleup}, such correlations would be entirely erased. However, at least some degree of these correlations can be preserved by orienting the halo-centric coordinates of the satellites along the direction of some larger-scale vector.

For example, suppose that the position of satellites in a halo exhibit correlations with $\vec{v}_{\mathcal{I}},$ the principal eigenvector of the inertia tensor of the density field of the parent halo. And let $\theta_{\rm i}$ refer to the angle between $\vec{v}_{\mathcal{I}}$ and the radial vector of the halo's $i^{\rm th}$ satellite, $\hat{r}_{\rm i},$ so that we have ${\rm cos}(\theta_{\rm i})\equiv\langle\hat{v}_{\mathcal{I}}\vert\hat{r}_{\rm i}\rangle.$ Then if we compute $\vec{v}_{\mathcal{I}}$ for each halo in the target simulation, after mapping satellites into their target halo it is straightforward to rotate each target halo's collection of satellites so that the angles ${\rm cos}(\theta_{\rm i})$ are preserved; note that this rotation exactly preserves the radial distribution of satellites within their host, as well as the relative distance between satellites in the halo. Creating mocks with this effect turned on and off allows for targeted investigation of the impact of satellite anisotropy on cosmological observables.

\subsection{Intrinsic alignments}
\label{subsubsec:intrinsicalignments}

Correlations between the intrinsic orientation of galaxies and the large-scale density field is an important potential source of systematic error that must be controlled for in lensing-based cosmological inference \citep{hirata_etal07,joachimi_etal11,mandelbaum_etal11c,heymans_etal13,chisari_dvorkin13,tenneti_etal15,troxel_ishak15}. Mock catalogs with variable levels of such intrinsic alignments (IA) have come to play a critical role in studies of this important systematic, and {\gsam} techniques can be extended to address this need.

The mathematics of generating mocks with intrinsic alignments is identical to the treatment of satellite anisotropies discussed in \S\ref{subsubsec:anisotropy}, and is closely related to the halo-based model introduced in \citet{schneider_bridle10}. We begin by associating some generic orientation vector $\vec{v}_{\rm halo}$ with every halo in both the source and target simulation; a conventional choice for $\vec{v}_{\rm halo}$ would be the halo shape vector or angular momentum vector for centrals and the radial position vector for satellites \citep[e.g.,][]{heavens_etal00, knebe_etal10, joachimi_etal11}, but in principle $\vec{v}_{\mathcal{I}}$ or some alternative choice could be used instead. If each galaxy in the source catalog has an orientation vector $\vec{v}_{\rm gal},$ then we can define an angle $\theta_{\rm align}$ with respect to $\vec{v}_{\rm halo}$ such that ${\rm cos}(\theta_{\rm align})\equiv\langle\hat{v}_{\mathcal{I}}\vert\hat{v}_{\rm gal}\rangle.$ Preserving these angles is rather simple: after mapping each galaxy from the source catalog into its host halo in the target simulation, the galaxy orientation vector $\vec{v}_{\rm gal}$ is rotated by an angle ${\rm cos}(\theta_{\rm align})$ with respect to the vector $\vec{v}_{\rm halo}$ of the target halo.

By construction, producing a mock catalog in this fashion perfectly preserves the distribution $P(\theta_{\rm align}\vert\mhalo)$ in the source catalog. Moreover, this technique will introduce an IA signal on cosmological distance scales when implemented with respect to any property $\vec{v}_{\rm halo}$ that exhibits large-scale correlations \citep[e.g., halo shape, triaxility, etc.,][]{schneider_etal12c}. In addition to the {\gsam} code responsible for the halo-to-halo correspondence, the {\tt matrix\_operations\_3d} module in {\tt Halotools} contains implementations of all the required three-dimensional rotations and dot products, and so publicly available software exists to carry out all the associated calculations with this approach to generating mocks with tunable levels of intrinsic alignments.

\subsection{Multi-simulation sampling}
\label{subsec:multisim}

In the mock-making program described in \S\ref{sec:scaleup}, there exists a single catalog of source galaxies that is transferred to the larger-volume target halo catalog. However, there is no strict need for the source of baseline galaxies to be unique: the framework outlined here naturally supports multiple sources of galaxies/halos. For example, suppose one wishes to construct a synthetic sky generated from two distinct hydrodynamical simulations: one higher-resolution run in a smaller box, and another large-volume, lower-resolution simulation that contains a larger population of cluster-mass halos. When constructing the halo-to-halo correspondence, for every target halo one simply searches the concatenated collection of simulated source halos for a close match. We are currently using this variation on our methods to generate mock catalogs designed to study the impact of systematic errors associated with baryonic effects that are not present in the underlying $N$-body simulation.

\begin{figure}
\centering
\includegraphics[width=8cm]{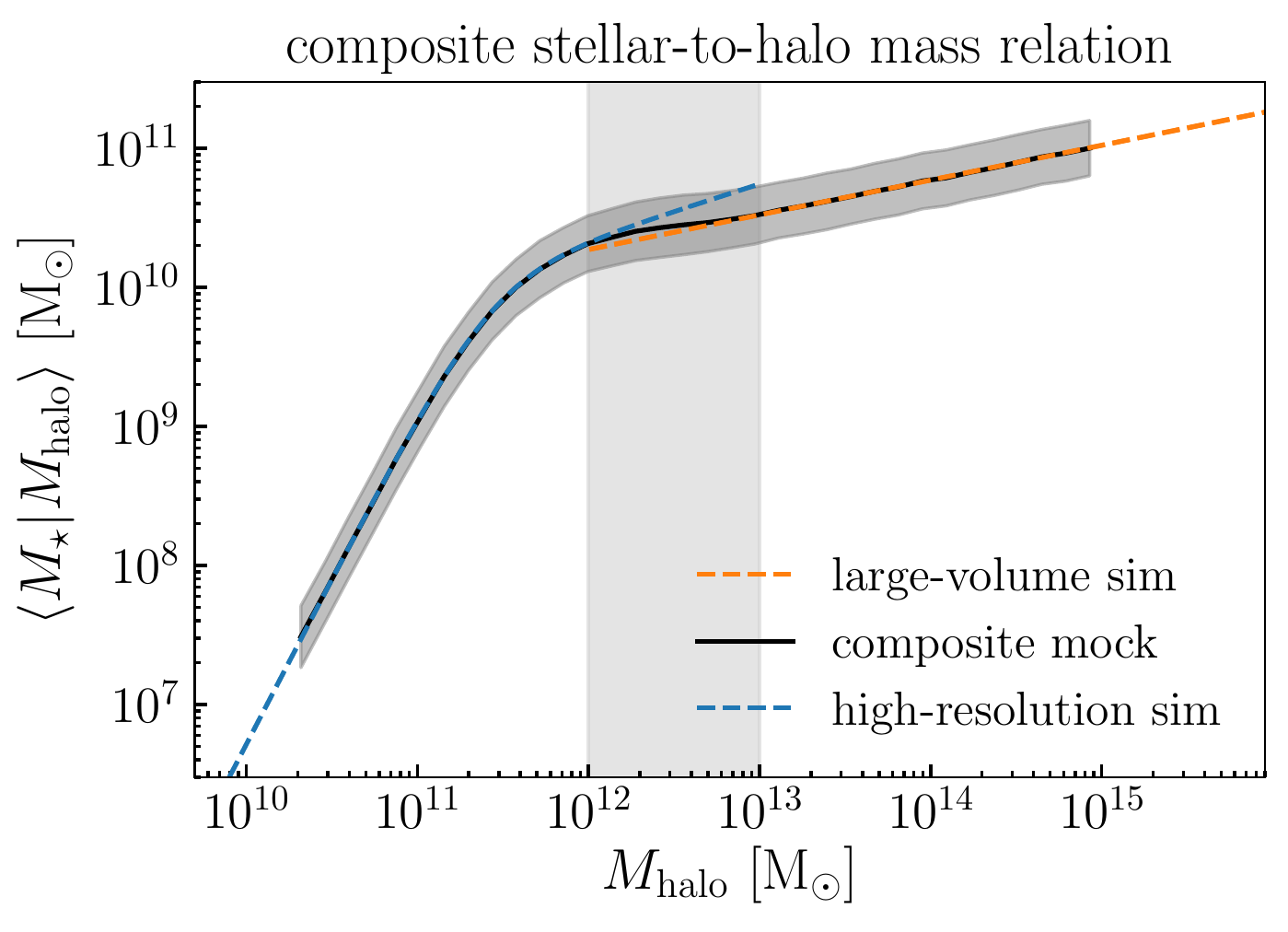}
\caption{
Toy example illustrating the technique outlined in \S\ref{subsec:multisim}. The stellar-to-halo-mass relation is compiled from two distinct simulations that span different halo masses due to differences in volume and resolution. The dashed orange curve shows the approximate power-law relation in the larger-volume, lower-resolution simulation; the dashed blue curve shows the same for the smaller-volume, higher-resolution simulation. We construct a composite relation $\langle M_{\star}\vert M_{\rm halo}\rangle$ via weighted-sampling the two simulations in the overlapping mass range shown with the gray vertical band, allowing us to populate a large-volume halo catalog with a scaling relation that derives from the two distinct baseline simulations.}
\label{fig:multisim_smhm}
\end{figure}

An important effect to account for in this multi-simulation application is to smooth out any ``seam" that may be present in distribution of source halo masses. Figure \ref{fig:multisim_smhm} uses a toy example to illustrate a simple technique for this smoothing procedure. We have generated two samples of synthetic galaxy data that serve as source galaxies/halos in this toy example. The first sample of halos spans the range $10^{10.25}\lesssim\log M_{\rm halo}/M_{\odot}\lesssim10^{13},$ the second spans $12<\log M_{\rm halo}/M_{\odot}<15.$ In each sample of synthetic halos, we have used the stellar-to-halo-mass relation taken from \citet{moster_etal13} to populate halos with synthetic galaxies with mass $\mstar,$ using different model parameters for the two halo samples.

We perform a weighted selection of the two samples by probabilistically selecting galaxies according to a sigmoid function:
$$
P_{\rm select}(\mhalo) = p_{\rm min} + \frac{p_{\rm max}-p_{\rm min}}{1 + \exp(-k\cdot(\log\mhalo - \log M_0))},
$$
where $\log M_0=12.5,$ $k=5,$ and $(p_{\rm min}, p_{\rm max})=(0, 1)$ for the large-volume simulation, and $(p_{\rm min}, p_{\rm max})=(1, 0)$ for the high-resolution simulation. The black curve in Figure \ref{fig:multisim_smhm} shows the stellar-to-halo mass relation in the composite mock catalog defined by the weighted sampling.

\section{Discussion}
\label{sec:comparison}

As described in \S\ref{sec:fundamentals}, a core assumption of {\gsam} is that the galaxy-halo connection is fundamentally sparse. Many alternative techniques for generating synthetic cosmological data rely upon this assumption. For example, the ADDGALS model (Wechsler et al. 2019, in prep) used in the Dark Energy Survey\footnote{\url{https://www.darkenergysurvey.org}} \citep[DES,][]{maccrann_etal18,buzzard19} is trained on an abundance matching mock, and additional galaxy properties beyond luminosity are drawn directly from observational datasets. Similarly, the basis of the MICE model \citep{carretero_etal15} used in DES and Euclid is a variation of the color-dependent HOD model developed in \citet{skibba_sheth11}; in the MICE mocks, tertiary galaxy properties beyond $M_{\rm r}$ and $\gr$ passively inherit their connection to the underlying density field, as described \S\ref{subsec:rescaling}.

By contrast, mocks based directly on {\tt Galacticus} \citep{benson_galaxy_2012} are being used by the WFIRST survey \citep{spergel_etal15}. Strictly speaking, semi-analytic models such as {\tt Galacticus} also make an assumption of the form shown in Eq.~\ref{eq:fundamental_decomposition}. In particular, in {\tt Galacticus} and most semi-analytic models, galaxy properties are closely connected to halo merger history, which itself exhibits dependence upon the density field over a wide range of spatial scales. Thus in SAMs, $P(\gvec\vert\delta)$ is not sparse in the sense that a large number of galaxy properties exhibit a complex high-dimensional connection to the density field.

{\gsam} is a fundamentally halo-based methodology, and so the techniques described here critically depend upon the ability to resolve host halos in the target simulation. This differentiates {\gsam} from methods based on approximate $N$-body simulations \citep[see][and references therein]{chuang_etal15, monaco16}, which are generally far less computationally demanding of the target simulation. The halo-based nature of our methodology also contrasts with the ADDGALS technique. In ADDGALS, a baseline catalog generated with abundance matching is used to constrain a model for the distribution of the dark matter density field, $P(\delta\vert M_{\rm r}, z),$ allowing ADDGALS to populate a large-volume simulation that need not resolve all halos hosting the galaxies in the mock. The simulation demands of {\gsam} are comparable to those required by the MICE mocks used in DES \citep{carretero_etal15}, which are based on HOD-type techniques and therefore require the target simulation to at least resolve the host halos of the galaxies of interest. {\gsam} does not require the target simulation to resolve subhalos or merger trees, and so our methods are less computationally demanding relative to attempts to directly populate the target simulation with a semi-analytic model.

Monte Carlo sampling from a SAM library is complementary to alternative techniques that draw directly from subsampled data, and/or draw from SED templates. In a {\gsam} mock, each galaxy represents a physical solution to the system of equations modeled by the SAM. The utility of the semi-analytic model is especially clear in cases where it is necessary to extrapolate beyond the reach of existing data; such extrapolations are unavoidable when designing mocks for future surveys such as LSST, and SAMs offer a physically motivated way to generate mock galaxies in regimes where there is little or no data.

By recasting semi-analytic models as galaxy libraries, we place different demands on the semi-analytic model relative to conventional evaluation criteria. For example, when fitting the parameters of a SAM, one might require the mean color $\langle \gr\vert M_{\star}\rangle$ to scale properly with stellar mass. In the ideal case, the training of the original SAM would have been carried out in a simulation of the same cosmology as the target simulation, such that the scaling relation $\langle \gr\vert M_{\star}\rangle$ in the original SAM already agrees with the validation criteria within the required tolerance. However, {\gsam} does not necessarily prize SAMs with high-accuracy mean relations; instead, what is needed is an underlying library that {\em densely samples} and {\em spans the range} of the core galaxy properties that drive the validation criteria. With such a library, it is guaranteed that the resampled SAM galaxies will have statistical distributions that are in accord with the (hopefully) well-trained empirical model.

\section{Conclusion and Outlook}
\label{sec:conclusion}

The field of large-scale structure cosmology has become driven by galaxy surveys whose analysis requires extensive use of highly complex, ${\rm Gpc}$-scale synthetic galaxy catalogs. As new measurements become available from ongoing surveys such as DES, KiDS\footnote{http://kids.strw.leidenuniv.nl}, and HSC\footnote{\url{http://www.subarutelescope.org/Projects/HSC}}, near-future surveys such as LSST and the Dark Energy Spectroscopic Instrument \citep[DESI,][]{desi_design16} continually update the validation criteria used to assess the realism of mock data.\footnote{For example, the validation of mocks in LSST DESC is enforced through the continually updated DESCQA quality assurance software \citep{descqa}.} The evolving nature of the validation criteria of modern galaxy surveys necessitates using models that are not costly to refit; the simultaneous need for computational efficiency and physical complexity presents a challenge that will only continue to grow as galaxy surveys progress.

The weighted sampling methods described in \S\ref{sec:scaleup} and \S\ref{sec:matchup} give a flexible way to generate synthetic galaxy data that is subject to evolving constraints. Retuning the mock can be accomplished by adjusting the empirical model, speeding up training time by orders of magnitude beyond what is possible when fitting a traditional SAM; at the same time, mocks made with {\gsam} inherit realistic levels of complexity from the underlying semi-analytic model or hydro simulation. And as shown in \S\ref{sec:systematics}, halo-based sampling methods are also a natural way to create mocks that incorporate a wide variety of systematic effects. Our python code implementing these techniques is publicly available on GitHub at \url{https://github.com/LSSTDESC/galsampler} and \url{https://github.com/astropy/halotools}.

\section{Acknowledgements}

This paper has undergone internal review in the LSST Dark Energy Science Collaboration by Matt Becker, Joe DeRose, and Tom McClintock. We are grateful to Yao-Yuan Mao, Katrin Heitmann, and Salman Habib for useful discussions that benefitted this work at many stages. APH thanks Makaya McCraven for {\em In the Moment.}

This work was conducted in part at Aspen Center for Physics, which is supported by National Science Foundation grant PHY-1607611. Work done at Argonne National Laboratory was supported under the DOE contract DE-AC02-06CH11357. We gratefully acknowledge use of the Bebop cluster in the Laboratory Computing Resource Center at Argonne National Laboratory. The authors gratefully acknowledge the Gauss Centre for Supercomputing e.V. (www.gauss-centre.eu) and the Partnership for Advanced Supercomputing in Europe (PRACE, www.prace-ri.eu) for funding the MultiDark simulation project by providing computing time on the GCS Supercomputer SuperMUC at Leibniz Supercomputing Centre (LRZ, www.lrz.de). The Bolshoi simulations have been performed within the Bolshoi project of the University of California High-Performance AstroComputing Center (UC-HiPACC) and were run at the NASA Ames Research Center.

We thank the {\tt Astropy} developers for the package-template \citep{astropy}, as well as the developers of {\tt NumPy} \citep{numpy_ndarray}, {\tt SciPy} \citep{scipy}, Jupyter \citep{jupyter}, IPython \citep{ipython}, Matplotlib \citep{matplotlib}, and GitHub for their extremely useful free software. While writing this paper we made extensive use of the Astrophysics Data Service (ADS) and {\tt arXiv} preprint repository.

APH led development of {\gsam} and drafted the paper. DK and EVK extensively tested, extended, and helped debug the software used throughout {\gsam}, improving the code's capability for large-scale production runs. AB provided consultation and guidance at every stage, and closely supervised the application of {\gsam} to the {\tt Galacticus} library. HA developed the multi-simulation sampling code used to stitch different source catalogs together, the feature required by the method of incorporating baryonic effects discussed in\S\ref{subsec:multisim}. CB extended the Cython implementation of Conditional Abundance Matching in {\tt Halotools} to enable the CAM-based resampling discussed in \S\ref{subsec:rescaling}. DDC led the intrinsic alignment and satellite anisotropy systematics techniques discussed in \S\ref{subsubsec:intrinsicalignments} and \S\ref{subsubsec:anisotropy}. All co-authors directly contributed to the text and its revision.

The DESC acknowledges ongoing support from the Institute National de Physique Nucléaire et de Physique des Particules in France; the Science \& Technology Facilities Council in the United Kingdom; and the Department of Energy, the National Science Foundation, and the LSST Corporation in the United States. DESC uses the resources of the IN2P3 Computing Center (CC-IN2P3–Lyon/Villeurbanne - France) funded by the Centre National de la Recherche Scientifique; the National Energy Research Scientific Computing Center, a DOE Office of Science User Facility supported by the Office of Science of the U.S. Department of Energy under contract No. DE-AC02-05CH11231; STFC DiRAC HPC Facilities, funded by UK BIS National E-infrastructure capital grants; and the UK particle physics grid, supported by the GridPP Collaboration. This work was performed in part under DOE contract DE-AC02-76SF00515.

\bibliographystyle{mnras}
\bibliography{main.bib}

\end{document}